\documentclass[conference,10pt]{IEEEtran}
\usepackage{amssymb}
\usepackage{times}
\usepackage{amsmath}
\usepackage{graphicx}
\usepackage{subfigure}
\usepackage{algorithm}
\usepackage{algorithmic}
\usepackage{cite}
\usepackage{url}
\usepackage{multirow}
\usepackage{stfloats}
\usepackage{float}
\usepackage{epstopdf}

\begin{document}

\title{Prometheus: LT Codes Meet Cooperative Transmission in Cellular Networks}
\author{\IEEEauthorblockN{Hai Wang, Zhe Chen, Qingyuan Gong, Weidong Xu, Xu Zhang and Xin Wang}
\IEEEauthorblockA{School of Computer Science\\ Fudan University\\
\{11210240059,zhechen13,12210240057,12210240106,09300240019,xinw\}@fudan.edu.cn}
}
\maketitle
\begin{abstract}
Following fast growth of cellular networks, more users have drawn attention to the contradiction between dynamic user data traffic and static data plans. To address this important but largely unexplored issue, in this paper, we design a new data plan sharing system named Prometheus, which is based on the scenario that some smartphone users have surplus data traffic and are willing to help others download data. To realize this system, we first propose a mechanism that incorporates LT codes into UDP. It is robust to transmission errors and encourages more concurrent transmissions and forwardings. It also can be implemented easily with low implementation complexity. Then we design an incentive mechanism to choose assistant users (AUs), all participants will gain credits in return, which can be used to ask for future help when they need to download something. Finally real environment experiment is conducted and the results show that users in our Prometheus not only can manage their surplus data plan more efficiently, but also achieve a higher speed download rate.
\end{abstract}


\section{Introduction}

With the development of wireless network technology, mobile phones are rapidly becoming the central communication devices in our daily lives. Besides voice service, data service has turned into the focus of competition among service providers. For example, China mobile in the data traffic has carried on many explorations, included the motion dream network, the handset newspaper, 12530, 139 mailboxes, to fly letter, 12580, handset maps and so on.

%

However, we have found that today's mobile data service has a lot of shortcomings. It is probably best known that many ISPs just provide fixed data plans, overuse will cause rapid increase in cost and underusing will result in a lot of waste. Thus users have to pay attention to their data usage. What's more, as users' demand grows, downstream rate offered by current cellular networks is becoming increasingly insufficient. As a result, users often fail to get high quality of Internet service.

%

Thus, it is of great significance to figure out an effective mechanism to make use of idle wireless interfaces and improve cellular bandwidth without unacceptable expense. Borrowing some mobile phones to assist forwarding data is always practical. Cooperative transmission can achieve higher transactional performance, better server utilization, and lower delay, altogether adding up to potentially significant cost savings \cite{CTJTU}.

\begin{figure}
\begin{center}
\includegraphics[width=0.4\textwidth]{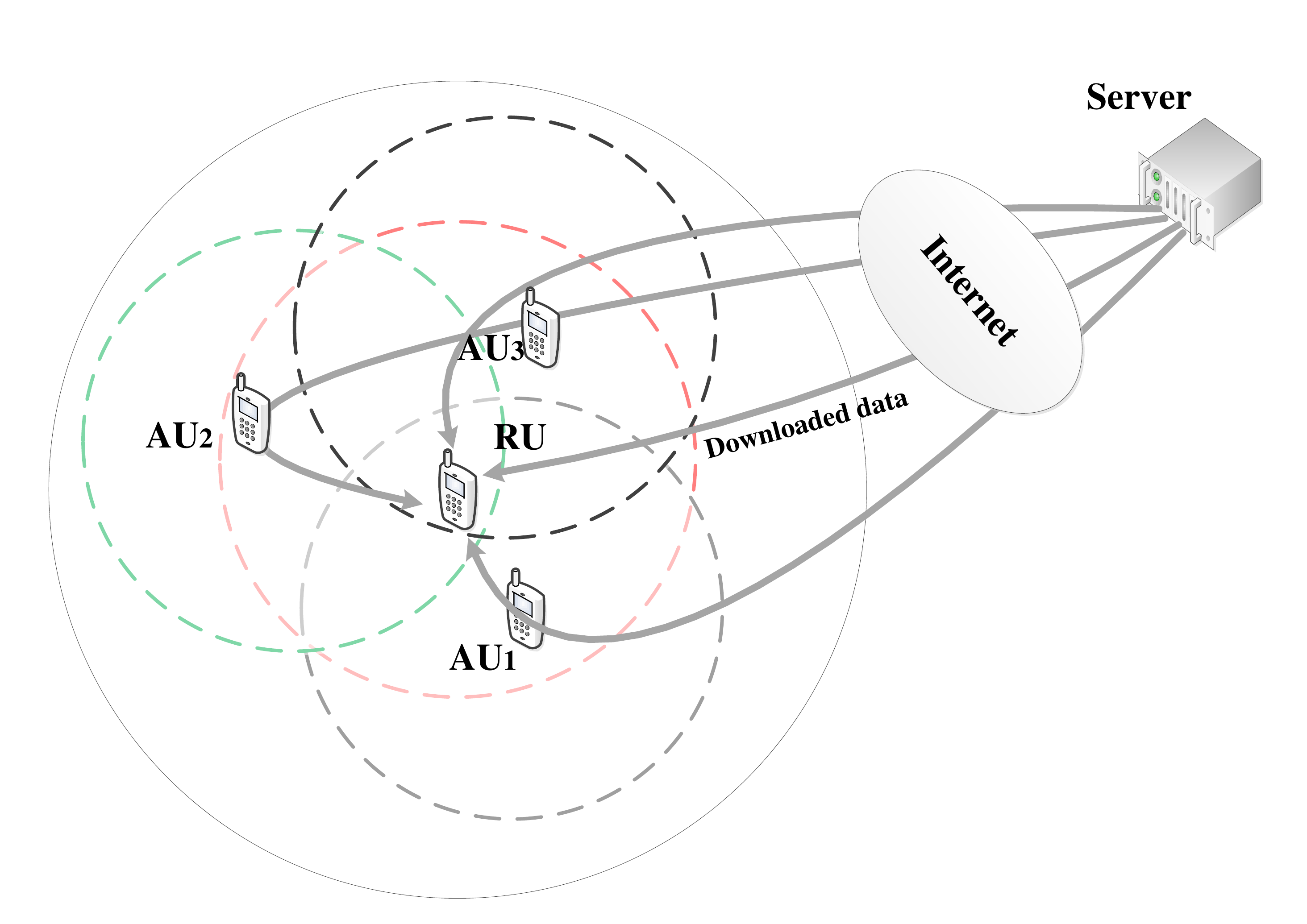}
\caption{Cooperative Transmission System: ${AU}_{1}$, ${AU}_{2}$, ${AU}_{3}$ help $RU$ to download something from server and then $RU$ could receive data from four distinct paths.}
\label{ct}
\end{center}
\end{figure}

In this paper, we construct a cooperative system. A set of users, termed as assistant users ($AUs$) between the server and the requesting user ($RU$) are first selected. Once the $AUs$ are determined, partial data packets originated from the server will be forwarded to the $RU$ by groups of $AUs$ relaying, the rest will be downloaded by $RU$ itself. As Fig.\ref{ct} shows, suppose $AU_1$, $AU_2$, $AU_3$ are ready to help $RU$. $AU_1$, $AU_2$, $AU_3$ download data from server, then $RU$ could receive data from four distinct paths. After the requested file data is integrated at $RU$, termination signal from server would be sent to the three partners. Everybody can celebrate that mission is over at much lower cost.

However, what is the apparently crucial path in cooperative transmission? It can be various wireless transmission mode of mobile device such as WLAN, Bluetooth and more. All of them are unstable and changeable over time. In the environment of multiple unstable mobile devices involved, there are two goals we must realize: 1) Indigenous drawback of multiple lossy links should be eliminated to make the cooperation as effective as possible. 2) Cooperative transmission could not be affected if some initial assistant users work halfway.

To achieve the two goals, we entail LT codes into the system. LT codes are forward error correction (FEC) codes and have emerged as an important potential approach to the operation of communication networks, especially wireless networks. The major benefit of LT codes stems from its ability to mix data, across time and across flows. This makes data transmission robust and effective over lossy wireless networks. The practical benefits of LT codes have been validated by many experiments.  It can not only ensure data reliable transmission, but improve wireless throughput substantially \cite{NCRedun,VANETNC1,VANETNC2}.

The main contributions of our work are shown as follows:

\begin{itemize}
\item We design and evaluate a cooperative transmission system that some smartphone users have surplus data traffic and are willing to help others download data.

\item To the best of our knowledge, this is the first attempt at combining LT codes with mobile cooperative transmission to improve the system performance.

\item The system has been implemented in Android platform and experiments present the performance of our system.
\end{itemize}

The rest of the paper explains the details of our new system along with its theoretical basis, and analyzes its performance using a practical system and simulations. In section \ref{sec2}, we present the related work. In section \ref{sec3}, we introduce the system model. The incentive mechanism for choosing AUs is designed in section \ref{sec4}. In section \ref{sec5}, we discuss the specific mechanisms we developed. We present performance results and an analysis in section \ref{sec5}. Finally we conclude the paper and present some future works.

\section{Related Work}\label{sec2}

 Cooperative transmission has been studied for many years. Most VANET protocols are designed based on the assumption of node cooperativeness \cite{VANETNC1}. Wireless sensor networks (WSNs) are another areas where cooperative transmission often occurs \cite{WSN1,WSN2,WSN3}.

%

Research in recent years focuses on the cellular networks. SHAKE was proposed in \cite{shake1999}, which lets some mobile hosts temporarily connect mutually and selects several kinds of paths to communicate with outside. It could ensure mobile hosts get larger transfer capacity. However, not only mobile hosts but also correspondent hosts on the internet require special software for SHAKE. So it is difficult to be popularized. In \cite{shake2002,shakeicc}, Konishi et al. had conducted extensive research into the issues raised. In SHAKE, mobile hosts which are connected with fast local link can use multiple wireless links owned by each host simultaneously to communicate with hosts on the internet. SHAKE can improve the data transfer speed.

Many researchers in this field have turned their research on streaming application. Instead of sending the stream independently to each mobile device, the server distributes the video packets among the devices, over a long-range wireless link using WLAN technology; the devices then exchange the received packets each other over short-range wireless links \cite{MobileCT1,MobileCT2}.

INDAPSON \cite{CTJTU} is a new incentive data plan sharing system, which allows users to better manage their data plans and makes it possible for unfamiliar users to cooperatively download data from the Internet. INDAPSON can make use of idle wireless interfaces and improve cellular bandwidth without unacceptable expense to update the infrastructures of ISPs. But it just supports some TCP applications and has no practical sense to UDP. However, TCP is not well suited for lossy links, which are generally more prevalent in wireless systems \cite{NCmeetsTCP}. The practical performance of INDAPSON may be poor. Fountain codes is a good choice to improve performance in lossy wireless network.

%

LT codes are the first practical realization of fountain codes \cite{ECCLT} and have been applied to cooperative transmission. R. Cao et al\cite{JSAC2} developed generalized decomposed LT codes for cooperative relay communications. The cooperative system based on LT codes can ensure communication reliability on both source-relay and relay-destination. This is the first time that LT codes meet cooperative transmission in cellular networks.


\section{System Model}\label{sec3}

\begin{table}\label{notation}
\centering
\caption{Summary of Notation}
\begin{tabular}{|c|c|}
\hline
Symbol & Description \\
\hline
$m$  &  Message size \\
\hline
$l$  &  Block size \\
\hline
$AUs$  &  Assistant users set \\
\hline
$AU_i$  &  Assistant user $i$  \\
\hline
$RU$  & Requesting user\\
\hline
$\epsilon_i$  &  Unit cost of $AU_i$ \\
\hline
$t_i$  &   Service time of $AU_i$ \\
\hline
$\mu_i$  &  Utility of $AU_i$ \\
\hline
$\mu$  &  Utility of Server \\
\hline
$R$  &  Reward\\
\hline
$\gamma$  &  System parameter  \\
\hline
$t_n$   &  Encoded block \\
\hline
$s_k$   &   Original block \\
\hline
$\Psi (d)$  & Robust soliton degree distribution of LT codes \\
\hline
$P(d)$  &  Idea soliton degree distribution of LT codes \\
\hline
\end{tabular}
\end{table}

Let there be $k$ feasible users participating in the cooperative transmission, denotes as $AUs=\{AU_i, i=1,2,\ldots,k\}$. In fact, the cooperative transmission can be divided into two stages: On the first stage, $S$ sends the coded symbols to $RU$ and $AU_i$, respectively. On the second stage, $AU_i$ transmits the downloaded symbols to $RU$.

%
Because TCP is a reliable, connection oriented protocol, there's no need to add erasure code in TCP. UDP is a minimal message-oriented transport layer protocol. It provides no guarantees to the upper layer protocol for message delivery and retains no state of UDP messages once sent. For this reason, LT codes can be used to enhance data availability and fault-tolerant of UDP. Table \uppercase\expandafter{\romannumeral1} lists frequently used notations.

\begin{figure}
\begin{center}
\includegraphics[width=0.45\textwidth]{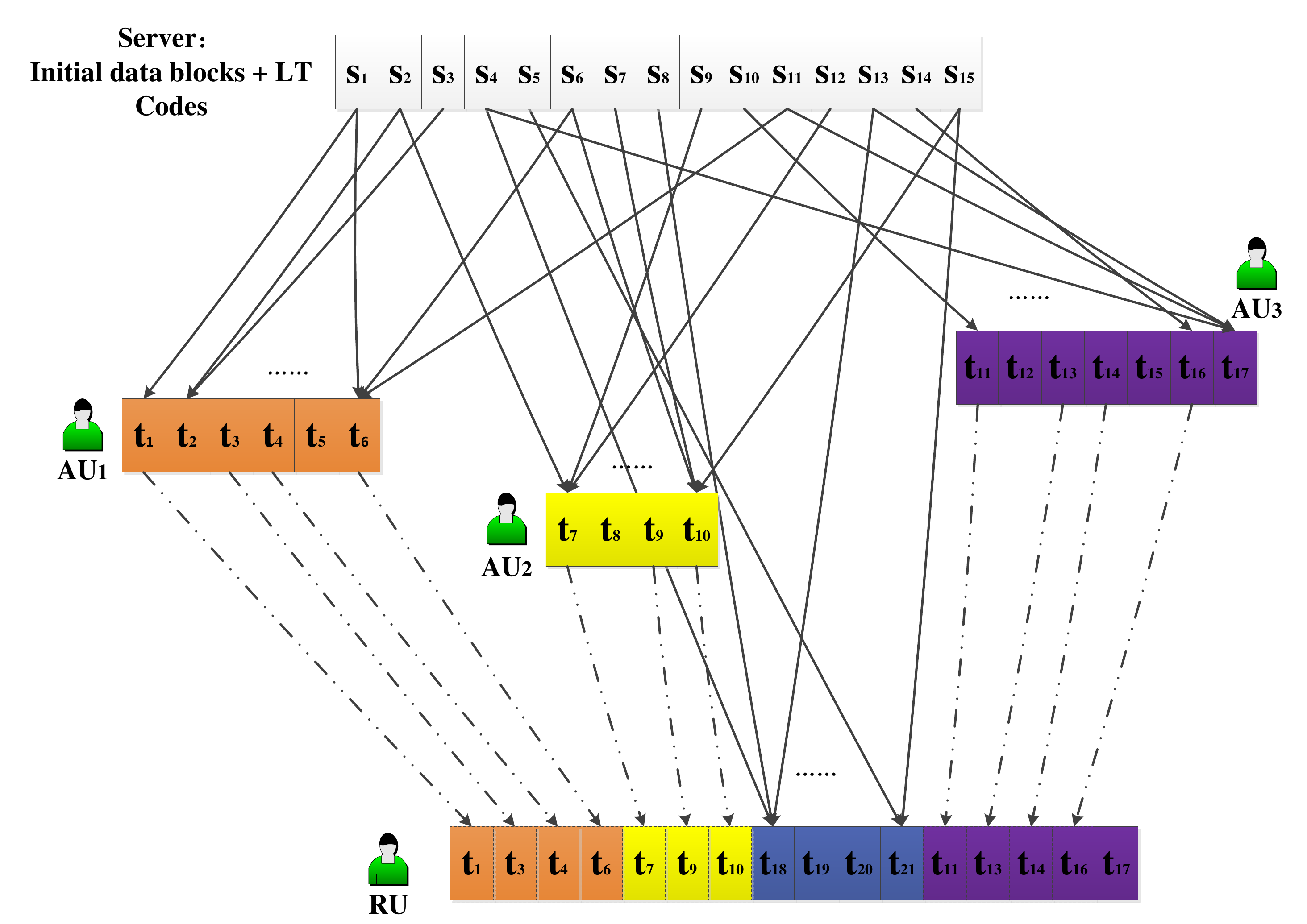}
\caption{LT codes meet cooperative transmission: Server encodes original symbols and sends different encoded packets to distinguishing clients.}
\label{ncudp}
\end{center}
\end{figure}

 Now consider server $S$ sending a message of size $m$ to requesting user $RU$. Assume that $S$ uses an erasure coding algorithm to generate code blocks of size $l$ such that $RU$ can decode the message after receiving any $(1+\epsilon)\frac{m}{l}$ code blocks \cite{ECCLT,LTcode}.
 Let $T$ denote the download time of $RU$. So the problem can be formulate as follows.

\subsection{Protocol Model}

\begin{figure}
\begin{center}
\includegraphics[width=0.5\textwidth]{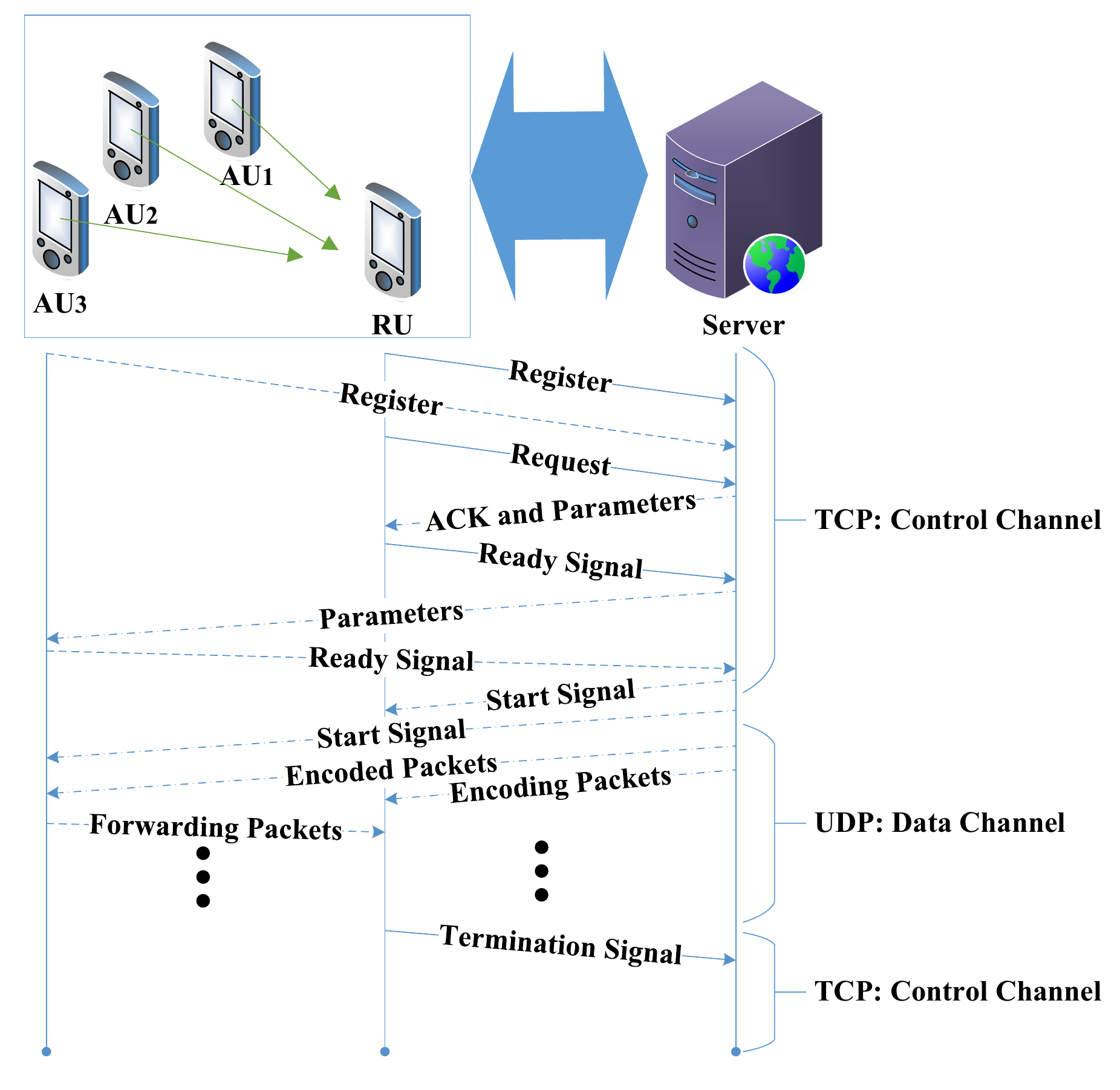}
\caption{System protocol model: All clients should register to server during period of time. $RU$ sends request to server when it need download service. Server finds $AUs$ nearby $RU$ according to clients' location information and sends relevant parameters to $AUs$. After that, cooperative download service is started. While $RU$ recoveries all original symbols, it tells server to terminate this cooperative download task.}
\label{protocol}
\end{center}
\end{figure}

 Fig.\ref{protocol} illustrates the workflow of Prometheus. Firstly, clients establish \emph{Control Channel}(TCP connection) to the server. Owing to network address translation(NAT) of base transceiver station(BTS), the server can't communicate with clients initiatively and then clients should register to server regularly. Client's registered information contains their IDs, Global Position System(GPS) information, power of smartphone, and so on. Secondly, if RU requests the server for help, the server generates an SSID randomly and sends it to RU and nearby AUs according to registered information. After that, RU establishes an soft access point(AP) automatically and AUs connect to it. If AUs connect to RU successfully, they will send a \emph{Ready Signal} to the server. Afterwards, they are grouped by the server. Thirdly, the server disseminates different encoded packets to clients using \emph{Data Channel}(UDP connectionless). AUs forward encoding packets to RU. As the file data are transmitted completely, RU sends a \emph{Terminating Signal} to the server and the cooperation process is done.

In practice, clients are dynamic. Considering some cases, Some AUs maybe leaving RU or the wireless link is broken. Even the wireless data rate is distinguished in varied time. If the dissemination strategy is static, some AUs finished forwarding but the others are busy. How to utilize all download links is significant for Prometheus. In \cite{CTJTU}, they use RU to divide download data into small segments as many tasks and assigns small tasks to AUs. Then AUs download these segments from resource server and forward to RU. With the help of LT codes, the fine-grained dissemination strategy is implemented in our system. The small segment is still large for Prometheus. The server only needs to disseminate encoded packets by multiple delivery paths. RU doesn't care about AUs' situations and it just concerns whether the received packets can recovery original data. We also evaluate our dissemination strategy in Subsection V-F.

\subsection{LT codes design}

Fig. \ref{ncudp} shows how we combine LT codes with cooperative transmission. Having run some tests, LT codes trade correction capabilities for computational complexity: practical algorithms can encode and decode with linear time complexity. LT codes can generate an arbitrary amount of redundancy symbols that can all be used for error correction. Receivers can start decoding after they have received slightly more than $M$ encoded symbols. However, there is only one parameter that can be used to optimize a straight LT code: the degree distribution function. Ideal soliton distribution is give as

\begin{align}\label{ideaLT}
 &P\{d=1\}=\frac{1}{n} \\
 &P\{d=j\}=\frac{1}{j(j-1)}, j=2,\ldots,n \nonumber
\end{align}

But the ideal soliton distribution does not work well in practice because any fluctuation around the expected behavior makes it likely that at some steps in the decoding process there will be no available packet of (reduced) degree $1$ so decoding will fail. Therefore, in our practice, a modified distribution, the "robust soliton distribution", is substituted for the ideal distribution, as shown in (\ref{robustLT}).

\begin{align}\label{robustLT}
 &\Psi\{d=j\}=\frac{(P(j)+\Theta(j))}{\beta}, j=1,2,3,\ldots,n
\end{align}
where $P(j)$ is the Ideal Soliton distribution and
\begin{align}\label{robustLT}
 \beta=\Sigma_{j=1}^{n}(P(j)+\Theta(j)) \nonumber
\end{align}

\begin{displaymath}
\Theta(j)= \left\{ \begin{array}{ll}
R/jn & \textrm{for $j=1,2,3,\ldots,n/(R-1)$}\\
R\ln(R\delta)/n & \textrm{for $j=n/R$}\\
0 & \textrm{for $j=n/(R+1),\ldots,n$}
\end{array} \right.
\end{displaymath}

\begin{displaymath}
\Theta(i)= \left\{ \begin{array}{ll}
\frac{R}{ik} & \textrm{for $i=1,2,3,\ldots,\frac{k}{R-1}$}\\
\frac{R\ln(\frac{R}{\delta})}{k} & \textrm{for $i=\frac{k}{R}$}\\
0 & \textrm{for $i=\frac{k}{R+1},\ldots,k$}
\end{array} \right.
\end{displaymath}

\begin{align}
 &R=c \sqrt{n} \ln(\frac{n}{\delta}), \textrm{for some suitable constant $c>0$} \nonumber
\end{align}

The effect of the modification is, generally, to produce more packets of very small degree (around $1$) and fewer packets of degree greater than $1$, except for a spike of packets at a fairly large quantity chosen to ensure that all original symbols will be included in some packets \cite{ECCLT}.  Algorithm \ref{ltcode} illustrates the detailed steps.

\begin{algorithm}
\caption{: LT Codes}
\label{ltcode}
\begin{algorithmic}[1]
\renewcommand{\algorithmicrequire}{\textbf{Encoding:}}
\renewcommand\algorithmicensure {\textbf{Decoding:} }
\REQUIRE ~~ \\
\STATE  Randomly sample the degree $d_n$ of the packet from degree distribution $\Psi (d)$.
\STATE  Choose, uniformly at random, $d_n$ distinct input symbols. Set $t_n$ equal to the bitwise sum, modulo $2$, of
these $d_n$ symbols. \\
\ENSURE ~~\\
\STATE  Find a packet $t_n$ which has exactly one source symbol $b_k$ in its list. If no such packet exists, the decoder
halts and fails. Otherwise: \\
  \quad (a) Set $b_k=t_n$. \\
  \quad (b) Set $t_{n^{'}}=t_{n^{'}}\oplus b_k$, for all packets $t_{n^{'}}$ which include source symbol $bs_k$ in their encoding lists. \\
  \quad (c) Delete source symbol $b_k$ from all encoding lists.
\STATE  Repeat step $3$ until all source symbols are decoded.
\end{algorithmic}
\end{algorithm}

\section{Incentive mechanism}

 In the real world, a user participating in the cooperative transmission will introduce overhead. Therefore their efforts will be rewarded. Taking cost and returning into consideration, each user makes its own cooperative plan. The chosen users will conduct the cooperative transmission. First they must download data from $S$ and then send the downloaded data to the requested user. This completes the whole multi-mode cooperative transmission process. After finishing the cooperative transmission, the $S$ (Server) computes the payment for each user and sends the payments to the users.

 Unfortunately, most of users are selfish by nature, they only want to maximize their own \emph{utility}. They will not participate in cooperative transmission unless there is sufficient incentive. At the same time, $S$ is also only interested in maximizing its own utility. To address this issue, we design an incentive mechanism using the \emph{Stackelberg game}. The mechanism consists of two stages: In the first stage, $S$ announces its reward $R$; in the second stage, each user strategizes sensing time to maximize its own utility. Both $S$ and the users are players. Therefore $S$ is the leader and the users are the followers \cite{gametheory}. The Stackelberg game can be solved to find the subgame perfect Nash equilibrium (SPNE), as follows.

\subsection{Model of Incentive mechanism}

 Meanwhile, $S$ announces a total reward $R > 0$, motivating users to participate in cooperative transmission, while each user decides its level of participation based on the reward. Let $t_i$ denote the number of time units $AU_i$ is willing to provide the retransmission service and $\epsilon_i$ is its unit cost. Then the utility of $AU_i$ is

\begin{equation}\label{AUiUtility}
\mu_i=\frac{t_i}{\sum_{j=1}^{k}t_j} R-t_{i}\epsilon_{i}
\end{equation}

The utility of $S$ is

\begin{equation}\label{RUUtility}
\mu=\gamma \log (1+\sum_{i=1}^{k}\log(1+t_i))-R
\end{equation}


where $\gamma > 1$ is a system parameter, the $\log(1 + t_i)$ term reflects the $S^{'}$s diminishing return on the work of $AU_i$, and the outer $\log$ term reflects the $S^{'}$s diminishing return on participating users \cite{MBS_Xue}.

Under this model, $S$ moves first and then the users move sequentially. $S$ considers what the best response of the follower is and then decide the optimal value of $R$ so as to maximize (\ref{RUUtility}). While each $AU_i$ actually observes this and in equilibrium picks the expected $t_i$ as a response to maximize (\ref{AUiUtility}). Since no rational user is willing to provide service for a negative utility, $AU_i$ shall set $t_i = 0$ when $R\leq \epsilon_{i}\sum_{j=1,j\neq i}^{k}t_j$.

\subsection{Determination of cooperative strategy}

To calculate the SPNE, the best response strategies of the users must first be calculated. Assume $S$ has announced a total reward $R > 0$, the next step is to determine $t_i$. Compute the first and second derivatives of $u_i$ with respect to $t_i$,

\begin{equation}\label{AUi1F}
\frac{\partial \mu_i}{\partial t_i}=\frac{-Rt_i}{(\sum_{j=1}^{k}t_j)^2} +\frac{R}{\sum_{j=1}^{k}t_j}-\epsilon_{i}
\end{equation}

\begin{align}\label{AUi2F}
&\frac{\partial^2 \mu_i}{\partial t_{i}^{2}}=\frac{2Rt_i(\sum_{j=1}^{k}t_j)-R(\sum_{j=1}^{k}t_j)^2}{(\sum_{j=1}^{k}t_j)^4}-\frac{R}{(\sum_{j=0}^{k}t_j)^2} \nonumber\\
& \qquad=-\frac{2R\sum_{j=1,j\neq i}^{k}t_j}{(\sum_{j=1}^{k}t_j)^3}
\end{align}

Since $\frac{\partial^2 \mu_i}{\partial t_{i}^{2}}<0$, the utility $u_i$ with respect to $t_i$ is a strictly concave function in. Therefore given any $R > 0$ and any strategy profile of the other $AUs$, the best response strategy of $AU_i$ is unique, if it exists. Setting the first derivative of $u_i$ to $0$ for maximisation, we have

\begin{equation}\label{AUiti}
t_i=\sqrt{\frac{R\sum_{j=1,j\neq i}^{k}t_j}{\epsilon_{i}}}-\sum_{j=1,j\neq i}^{k}t_j
\end{equation}


If the right hand side of (\ref{AUiti}) is positive, $t_i$ is the best response strategy of $AU_i$. If the subtraction result is less than or equal to $0$, $AU_i$ does not participate in the cooperative transmission and we set $t_i = 0$. Let $t=(t_1, t_2,\ldots, t_k)$ denote the strategy profile consisting of all users' strategies. Hence we have the following algorithm to compute $t$ \cite{MBS_Xue}.

\begin{algorithm}[H]
\caption{: Computation of $t$}
\label{NE}
\begin{algorithmic}[1]
\STATE Sort $AUs$ according to their unit costs, $\epsilon_{1}\leq \epsilon_{2}\leq \epsilon_{3}\leq \ldots \leq \epsilon_{k}$;\
\STATE $K\leftarrow \{1,2\},i\leftarrow 3$;\
\WHILE {$i < k$ and $\epsilon_{i}<\frac{\epsilon_{i}+\sum_{j\in K}\epsilon_{i}}{|K|}$}
\STATE $K\leftarrow K\cup \{i\}, i\leftarrow i+1$;
\ENDWHILE
\FOR{$i\leftarrow 0$ to $k$}
\IF{$i\in K$}
\STATE $t_i=\frac{(|K|-1)R}{\sum_{j\in K}\epsilon_{j}} (1-\frac{(|K|-1)\epsilon_{i}}{\sum_{j\in K}\epsilon_{j}})$;
\ELSE
\STATE $t_i=0$
\ENDIF
\ENDFOR
\RETURN $t=(t_1,t_2,\ldots, t_k)$
\end{algorithmic}
\label{STNE}
\end{algorithm}

The strategy profile $t = (t_1,t_2,\ldots,t_k)$ computed by Algorithm \ref{STNE} is an Nash Equilibrium (NE) of the Stackelberg game. Its time complexity is $O(n \log n)$. The whole proof process is beyond the scope of this paper and I will not repeat it now. If you are interested, please see \cite{MBS_Xue}.

\subsection{Server Utility Maximization}

Now the best response function of $S$ is considered. The best response is to find the value of $R$ that maximises $\mu$ given $t = (t_1,t_2,\ldots,t_k)$. Substituting $t_i$ into Equation (\ref{RUUtility}) and considering $t_i = 0$ if $i\notin K$, we have

\begin{equation}\label{U0MAX}
\mu=\gamma \log (1+\sum_{i=1}^{k}\log(1+T_{i}R))-R
\end{equation}

where

\begin{equation}\label{TiDef}
 T_i=\frac{(|K|-1)}{\sum_{j\in K}\epsilon_{j}} (1-\frac{(|K|-1)\epsilon_{i}}{\sum_{j\in K}\epsilon_{j}})
\end{equation}

For simplicity, let $X=1+\sum_{i=1}^{k}\log(1+T_{i}R)$. Then, differentiate $\mu$ with respect to $R$:

\begin{equation}\label{uF1}
\frac{\partial \mu}{\partial R}=\gamma \frac{\sum_{i\in K}\frac{T_i}{1+T_{i}R}}{X}-1,
\end{equation}

Based on the Equation (\ref{uF1}), we can easily get the second order derivative of $\mu$, which is always less than $0$. So there exists a unique maximizer $R^{\star}$ over $R\in [0,\infty)$. Now solving (\ref{uF1}) yields $R^{\star}$.  The bisection's method is used which possesses a good convergency, fast solution speed and high precision.


Over all, once we get $\mu$, we know how much $RU$ should pay for the cooperative transmission. It is shown below, which is hard to be lowered.

\begin{equation}\label{PayRU}
P_{RU}=\gamma \log (1+\sum_{i=1}^{k}\log(1+T_{i}R))-\mu_0
\end{equation}

\section{Design and Implementation}\label{sec4}
 The system is built on traditional Client/Server model.  This section provides a high-level overview of each implementation framework. In order to consider compatibility in different platforms, our Prometheus is done in Java language.

\begin{figure*}
\begin{center}
\includegraphics[width=0.85\textwidth]{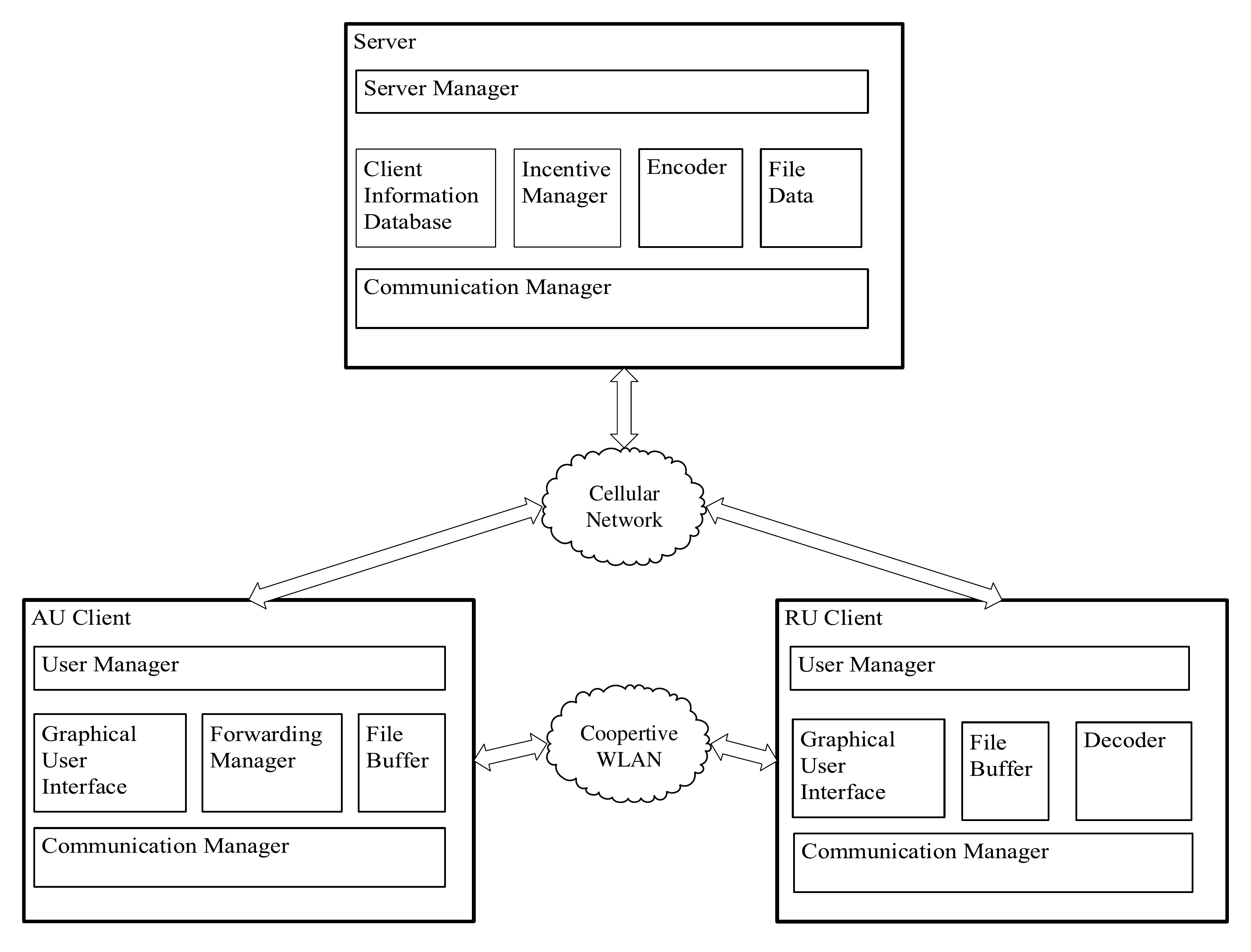}
\caption{A high-level overview of the implementation framework: That framework is a traditional Client/Server model. AU client and RU client are implemented on the smartphone.}
\label{server-a}
\end{center}
\end{figure*}

\subsection{Server Design}

Comparing to client, server is easy to implement. As shown in Fig.\ref{server-a}.

\emph{Server Manager} is a central controller which manages and communicates with other components.

\emph{Server Manager} not only manages Client Information Database and Incentive Manager, but also provides a graphical user interface and web manager for administrator of system.

\emph{Client Information Database} records the clients¡¯ information, such as IDs, credits and location informations.

\emph{Incentive Manager} gives an appropriate repayment to clients according to their own contribution.

\emph{File Data Storage} is used to store the data file which needs to be transmitted.

\emph{Encoder} receives original data from \emph{File Data Storage} and encodes them to LT coded symbols.

\emph{Communication Manager} controls wireless interfaces, such as WiFi interface and cellular interface. All network data packets are managed by \emph{Communication Manager}. \emph{Communication Manager} is responsible for distinguishing owners of data packets and exchanging them with upper components.

We implement the server side from basic building blocks: network socket programming, multithreading programming, LT coder, database programming, incentive manager, data packets processing and sending. When server receives a request from RU and  AUs have connected to it successfully, different encoded symbols are sent to clients by multithreading sockets. For the purpose of testing performance in varied packet loss rate, the network emulator NEWT(Network Emulator for Windows Toolkit) \cite{NEWT} is employed in \emph{Communication Manager}.

\subsection{Client Design}

Now we talk about the client side frameworks available to multiple concurrent versions of the Android OS. As shown in Fig.\ref{server-a}, there are two parts implemented in one Android application. The client only can choose PU or AU mode in our policy.

\emph{User Manager} is same as \emph{Server Manager} and every component is managed by it. \emph{User Manager} decides to enter either RU or AU mode based on user's action.

\emph{Graphical User Interface} provides an interface to users to control our Android Application. All the data file in our server will be displayed to users by GUI. Users can choose what they want to download.

\emph{Forwarding Manager} controls WiFi interface to forward received encoded packets to RU. It just works in AU mode.

\emph{File Buffer} is shared with AU and RU and all received encoded packets are stored here.

\emph{Decoder} is used to recovery original data and writes them in flash rom.

\emph{Communication Manager} is more complicated than server's because there are two modes in client. If RU, \emph{Communication Manager} turns on the soft access point(AP) mode for waiting AUs' forwarding data. At the same time, \emph{Communication Manager} will prevent AUs surfing Internet from RU. After \emph{Communication Manager} receives data, it uploads to \emph{Decoder} to recovery the original data. On the other hand, \emph{Communication Manager} controls \emph{Forwarding Manager} to forward received packets to RU, while user is AU.

In our daily life, smartphones are widespread and equipped with multiple wireless interfaces. AUs need to download data from server via 3G and forward them to RU via WiFi network. However, these multiple wireless interfaces are not permitted working simultaneously. We have to make some changes to enable it. Android phones are a good choice for our experiments because Android is an open-source project. The traditional Android phones also only allow WiFi network interface connection and turn off cellular network interface so as to save the power. This policy is maintained by Android Connectivity Service. We should break out the limitation and resort to the function startUsingNetworkFeature. This function will force Android phones connecting to Internet through cellular network in several seconds  when WiFi is turned on. Just like \cite{CTJTU}, The AUs can't pull data from Internet by RU without its permission.

The LT decoder is implemented in smartphones. Multiple received threads are created by RU to receive forwarding data from AUs. At the same time, LT decoder thread is created to decode encoded symbols from every AU. Obviously, This is a classical Producer-Consumer problem in multithreading programming design. All incoming encoded data flows are pushed to their corresponding FIFO queues and then LT decoder thread reads outgoing data flows from the FIFO queues. In order to evaluate goodput of Prometheus, the decoded data flows are counted by the monitor thread with a specific timeout value.

\section{Measurement and Evaluation}\label{sec5}

In this section, we test the effectiveness of Prometheus. The heterogeneous terminals should be considered, so we utilize different Android phones and operators. The type of cellular network is 3G, which has been widely used around the world. The detailed parameters is shown in Table \ref{ceshiPhones}. In order to better choose encoding parameters, such as the block size, symbol size and decoding overhead, LT codes' decoding throughput is measured on SAMSUNG I9300. In all experiments, unless specified otherwise, the block size means number of origin symbols. For the sake of comparison, we also implemented a conventional TCP cooperative transmission system, which has been carried out by \cite{CTJTU}.

To evaluate the performance of Prometheus, we evaluate some important metrics: (1) \emph{Goodput}: Goodput is the application layer throughput, i.e. the number of useful information bits delivered by the network to a certain destination per unit of time. It is measured on the RU side during cooperative transmission and can better reflect user's quality of service (QoS). (2) \emph{Decoding Overhead}: We measure the percentage of redundancy over original symbols to evaluate the level of redundancy when using LT codes. (3) \emph{Decoding Throughput}: We change block size and symbol size to measure the decoding throughput on SAMSUNG I9300. (4) \emph{Power Consumption}: Power consumption is one of the most important consideration to ensure the reliable and economical operation of cooperative system.

\begin{table*}
\centering
\caption{Summary of testing phones }
\begin{tabular}{|c|c|c|c|c|c|}
\hline
 Type &Quantity& CPU & RAM & Android Version& Mobile Network \\
\hline
Huawei T8951  &  1 & 1.0GHz dual-core &512MB & 4.0.2 &China Mobile (TD-SCDMA)\\
\hline
Motorola ME860  &  1 & 1.0GHz dual-core &1GB & 2.2.2 &China Unicom (WCDMA)\\
\hline
SAMSUNG I9220  & 1 & 1.4GHz dual-core &1GB & 4.0.3 &China Unicom (WCDMA)\\
\hline
SAMSUNG I9300  &  2 & 1.5GHz quad-core &1GB & 4.4.2 &China Unicom (WCDMA)\\
\hline
\end{tabular}
\label{ceshiPhones}
\end{table*}

\subsection{Goodput of Cooperative Transmission}

First we evaluate the download performance of Prometheus. We varied the number of AUs from 1 to 5 to test the average download rate. During the experiment all the smartphones had access to both 3G and WiFi connections. End-to-end message transmission at the transport layer can be categorized as either connection-oriented (TCP) or connectionless (UDP). In this subsection, experiments based on UDP and TCP are all proceeded.

\begin{figure}
\begin{center}
\includegraphics[width=0.45\textwidth]{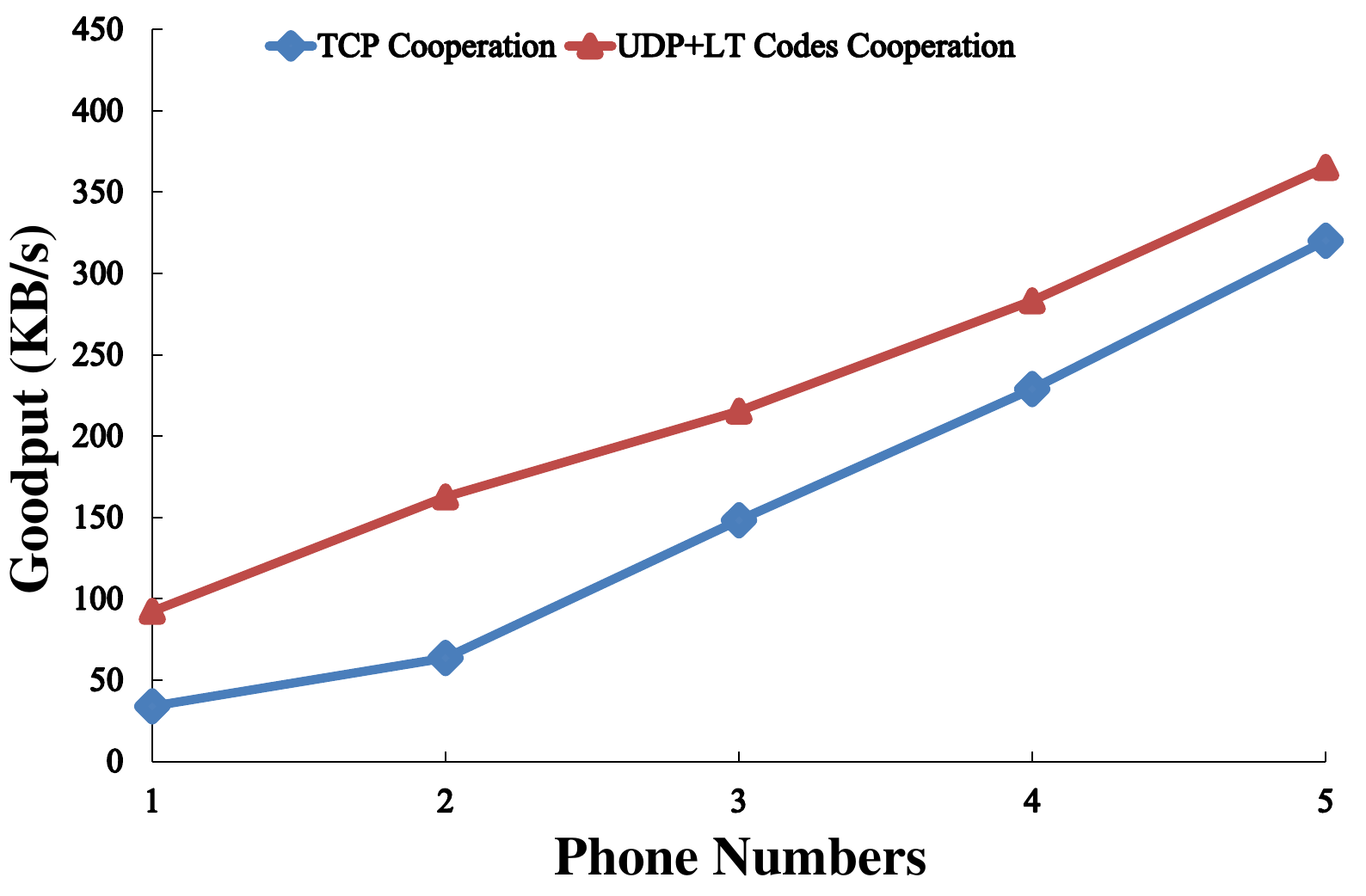}
\caption{The download rate of $RU$ with different number of phones: By comparing cooperative download based on TCP and Prometheus, our data plan sharing system's goodput is higher. The size of the download file is 46.75MB and every value shown in the figure is averaged over 10 experiments. }
\label{gvu}
\end{center}
\end{figure}

Fig.\ref{gvu} shows the average download rate of RU with different number of smartphones in the download group. It can be seen that the rate increases with the number of phones linearly. Since in general the data rate of WiFi is always larger than that of 3G, the download data rate for RU is proportional to the sum of the 3G download rates of all the phones. Although wireless signals are easily affected by environments, so the momentary rate of 3G changes with time, but the average 3G rate does not vary a lot. By comparing the average download rate with and without cooperation, we can see that taking the advantage of the idle 3G data interfaces of the AUs improves the download rate of the RU greatly. It is clear to see that cooperative transmission can linearly increase the download rate with constant average 3G data rate.

\subsection{Changing of AUs}

\begin{figure}
\begin{center}
\includegraphics[width=0.45\textwidth]{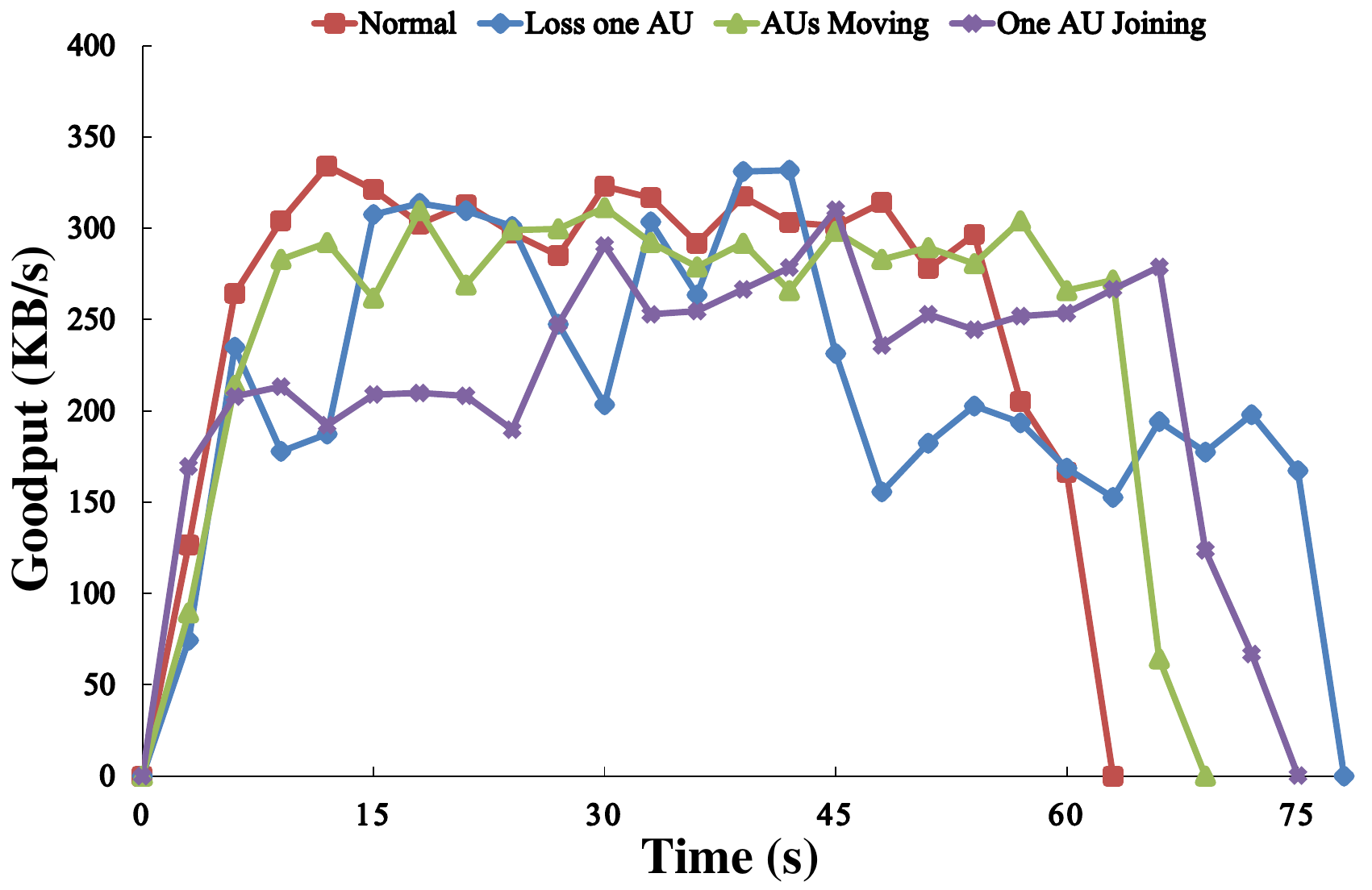}
\caption{Changing of $AUs$: There are three $AUs$ and one $RU$ to download a file with size 9.42MB in these experiments. No matter what $AUs$ change, they only impact on downloaded time.}
\label{cau}
\end{center}
\end{figure}

 Prometheus adopts strict dynamic management, which makes it easy to adapt to wireless communication environment with constant changes. When AUs join or leave the system, its only effect is to increase or decrease the download speed, and ultimately reduces or increase the time required for data to be downloaded from server. The mobility of AUs has even less influence on the service quality of system, as Fig.\ref{cau} shows. From what has been discussed above, we can safely draw the conclusion that the cooperative transmission system based on LT codes has good robustness and powerful ability to adapt changing network conditions.

\subsection{Effects of LT codes in Packet Loss}

\begin{figure}
\begin{center}
\includegraphics[width=0.45\textwidth]{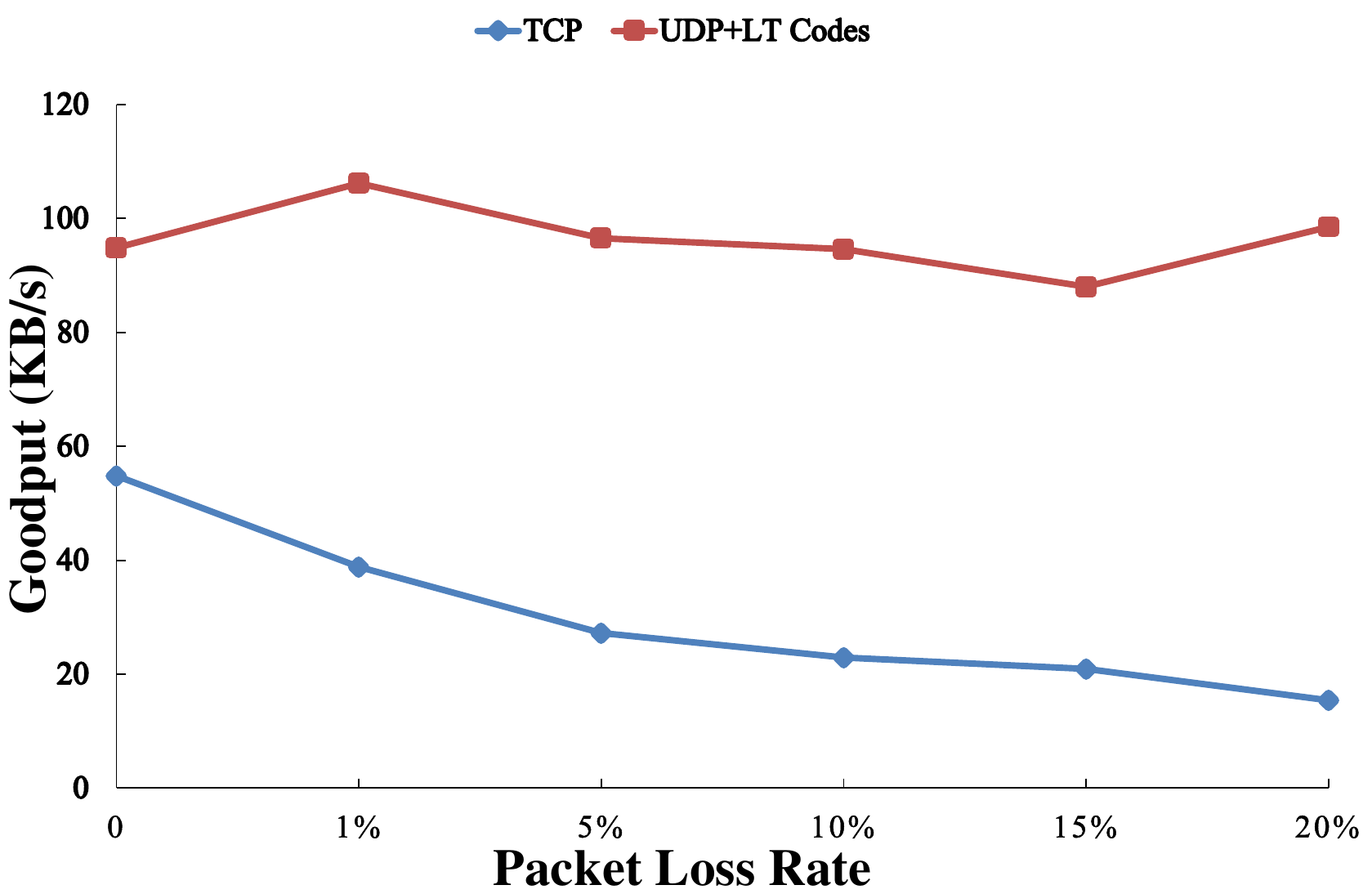}
\caption{One smartphone download data from server in different packet loss rate: Our data plan sharing system downloads file with size 9.42MB in only one smartphone without $AUs$. It shows better robust performance in the presence of packet loss than the smartphone who downloads file data based on TCP.}
\label{onelossrate}
\end{center}
\end{figure}

\begin{figure}
\begin{center}
\includegraphics[width=0.45\textwidth]{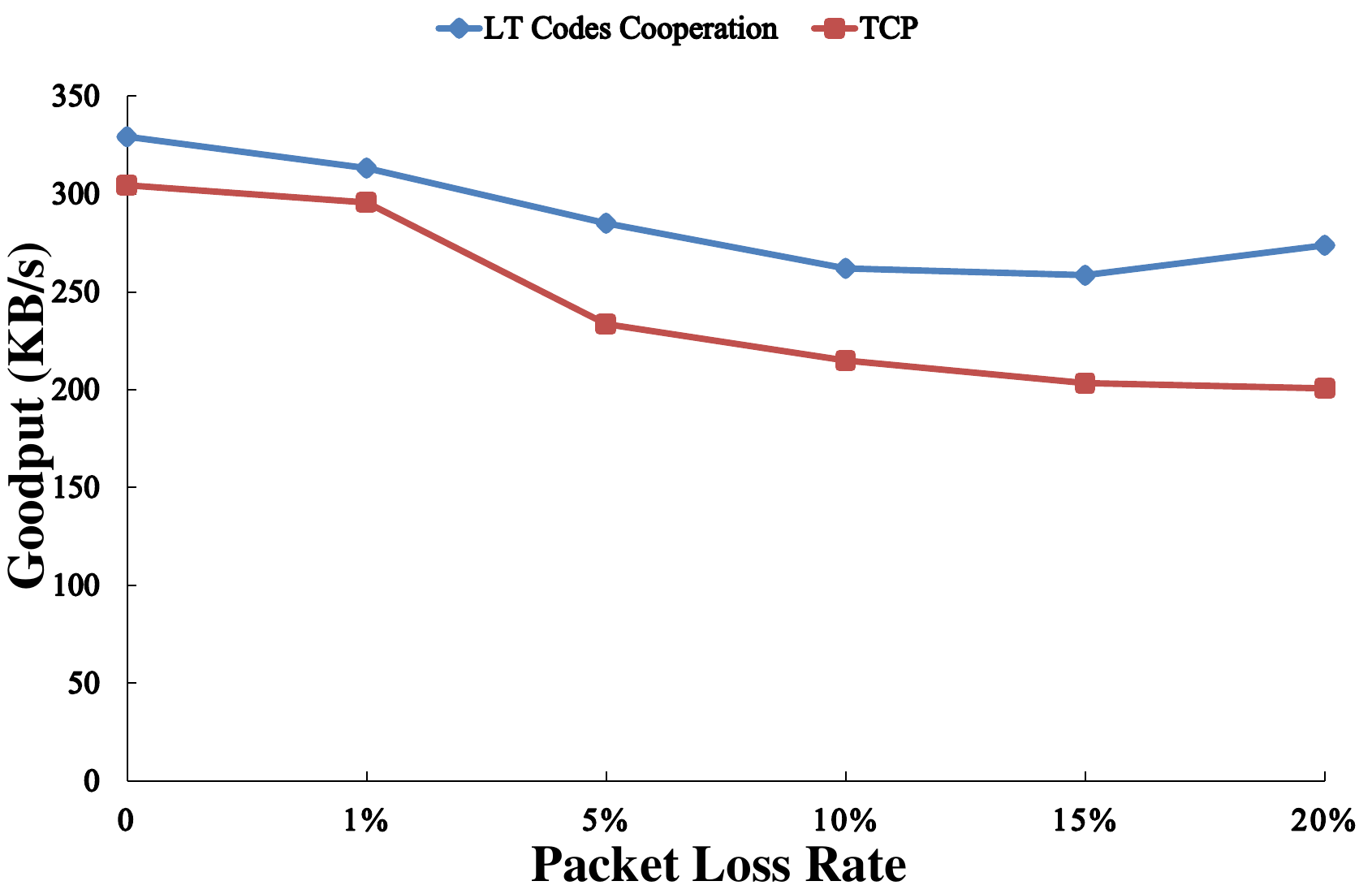}
\caption{Cooperative transmission are in different packet loss rate: Four $AUs$ help one $RU$ download data file with size 9.42MB in different packet loss rate. The figure also shows our data plan system has better robust performance to distinguishing packet loss rate.}
\label{mutilossrate}
\end{center}
\end{figure}

Packet loss is observed to be a familiar phenomenon with transmissions of wireless networks. Frequent undesired packet loss seriously degrades the network performance. At first, we randomly picked a user (ID = 3) and plot its performance in Fig.\ref{onelossrate}. Result indicates packet loss has greater impact on TCP. When packet loss rate increases, the goodput of TCP is down, while the goodput of UDP with LT codes is more stable with smaller fluctuation. This is because TCP is a reliable, connection oriented protocol, packet loss results in duplicate transmissions, which will cause network delay and can severely degrade performance.

Let $Ploss_{Max}$ denote the users' maximum range of packet loss rate. Fig.\ref{mutilossrate} shows the cooperative transmission performance decreases when $Ploss_{Max}$ ranges from 0 to 0.2. Although the difference between UDP with LT codes and TCP becomes smaller, Prometheus's goodput is still higher than the cooperative transmission based on TCP.

%

\subsection{Decoding Overhead}

Firstly, we evaluate the decoding overhead on SANSUNG I9300. When the communication between transmitter and receiver takes place in the Internet with Ethernet technology, there is the border of the maximum transmission unit(MTU). Most size of MTU in Ethernet LANs is 1500 bytes and if the larger MTU packet is pushed into next network layer, it's fragmented at the transmitter. Then we can do a simple estimation: Ethernet MTU - IP header(20 bytes) - TCP/UDP header(20/8 bytes) - application header $\approx$ 1450 bytes. So the max packet size in application layer is 1450 bytes.

\begin{figure}
\begin{center}
\includegraphics[width=0.45\textwidth]{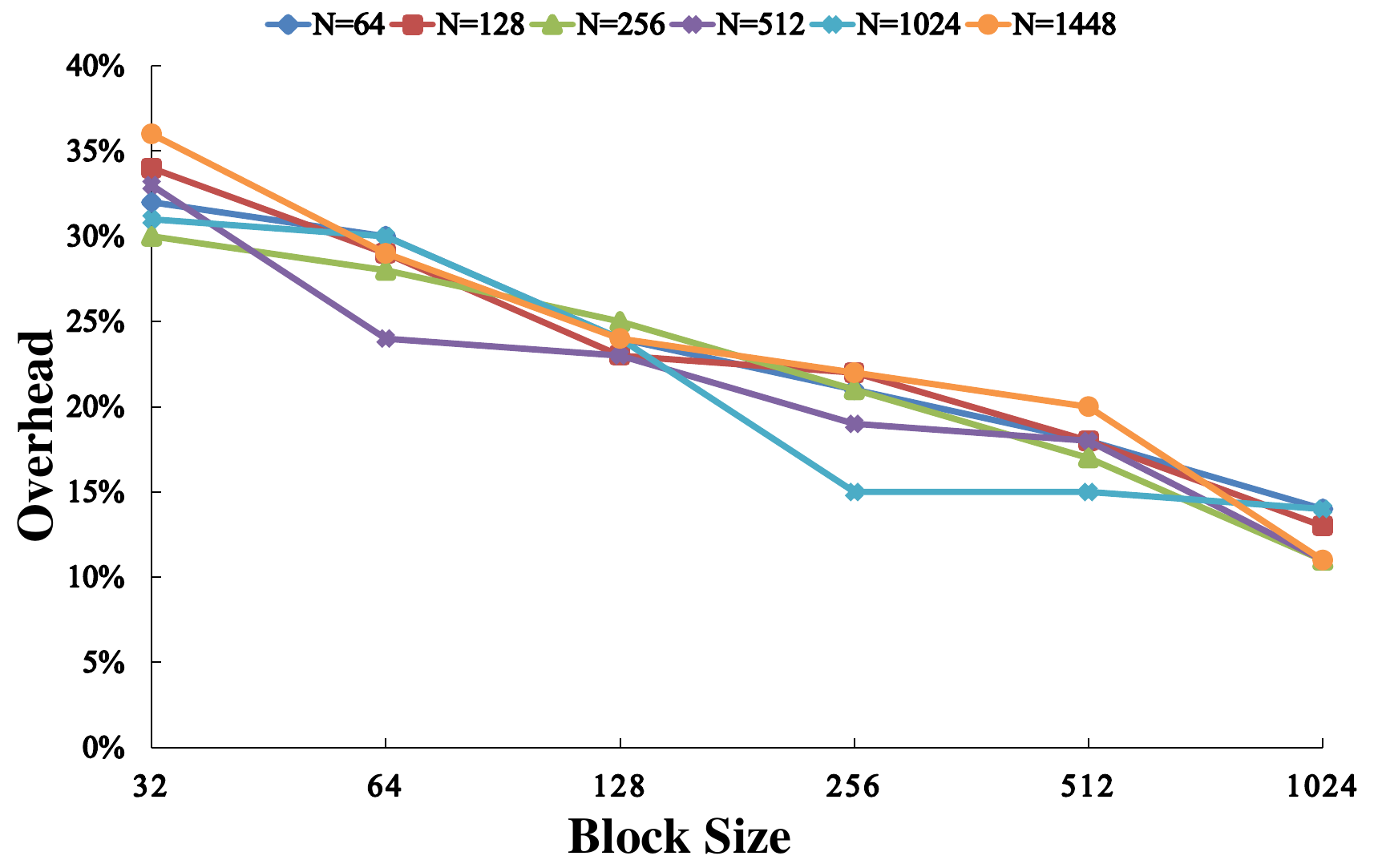}
\caption{Decoding overhead in SAMSUNG I9300}
\label{overhead}
\end{center}
\end{figure}

In order to determine how block sizes and symbol sizes impact on decoding overhead, we investigated a range of block size with k = 32, 64, 128, 256, 512, 1024 and a symbol size N = 64, 128, 256, 512, 1024, 1448 bytes. Then, a file is randomly cut into slices to encode in LT codes. We measure decoding overhead by every combination of block size and symbol size. Fig.\ref{overhead} shows that although N=512 at k=32 and N=1024 at k=256 are a little far away from other points, symbols size doesn't impact on decoding overhead heavily. For example, when symbol size is N = 1448 bytes, decoding overhead decreases from 36\% to 11\% with the increasing block size. Because more block sizes, more encoded symbols with one degree are generated by robust soliton distribution.

When we choose the block size in our Prometheus, tradeoff between block size and overhead rate needs to be considered. If the block size is large, more computation power and more memory space are needed. These resources are the most precious resources in our smartphones. When they are consumed much by LT decoder, smartphone will slow down severely or crash. In practice, buffer memory of each application is limited in Android device.

\subsection{Decoding Throughput}
\begin{figure}
\begin{center}
\includegraphics[width=0.45\textwidth]{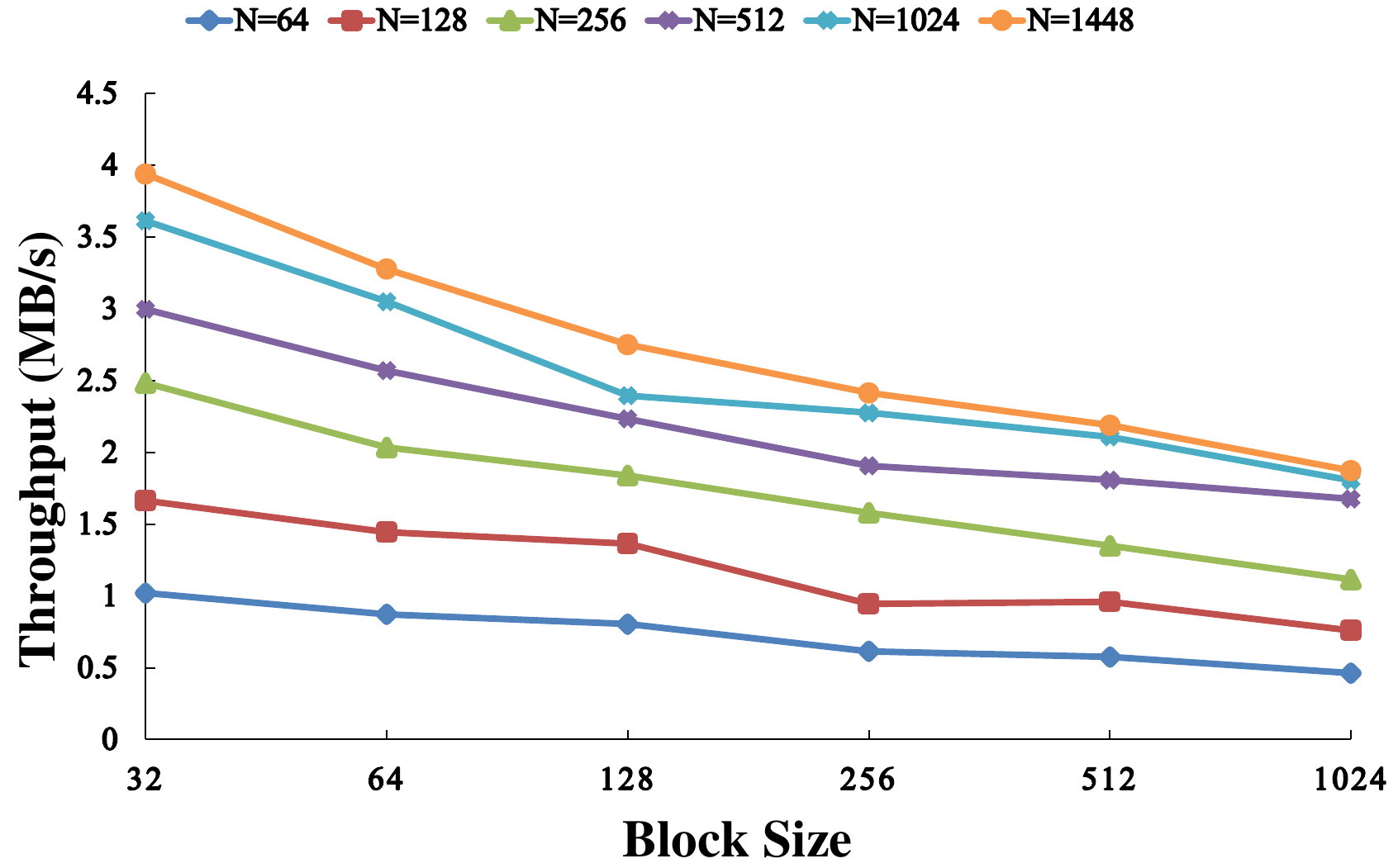}
\caption{Decoding throughput in SAMSUNG I9300}
\label{throughput}
\end{center}
\end{figure}
As an important metric to gain insights on Prometheus, we next examine decoding throughput on commodity smartphones. LT codes is simple and easy to be realized on smartphones. However, some important questions should be considered: Which size of symbol and block are chosen? What the maximum of decoding throughput can be reached on smartphones? We are interested in the influence of decoding throughput with different symbol size and block size. Obviously, the sum of all the AUs' goodput at RU must be less than its decoding throughput. We measure decoding throughput to choose the suitable symbol size and block size.

Decoding throughput is measured on SAMSUNG I9300 and we also investigate a range of block size with k = 32, 64, 128, 256, 512, 1024 and a symbol size N = 64, 128, 256, 512, 1024, 1448 bytes. It is easy to see in Fig.\ref{throughput} that the larger symbol size and smaller block size, the higher decoding throughput will be obtained. It seems k=32 and N = 1448 bytes can reach the maximum of decoding throughput but the small block size waste more repair symbols and pulls the decoding overhead rate up. For Prometheus, sending much data from server and making the cellular network busy will be bad for other mobile users. So as to balance decoding overhead and decoding throughput, we choose k=64 and N=1024 bytes in Prometheus and server will split up data file into 64KB segment for transmitting.

\subsection{Decoding Power Consumption}

\begin{figure}
\begin{center}
\includegraphics[width=0.45\textwidth]{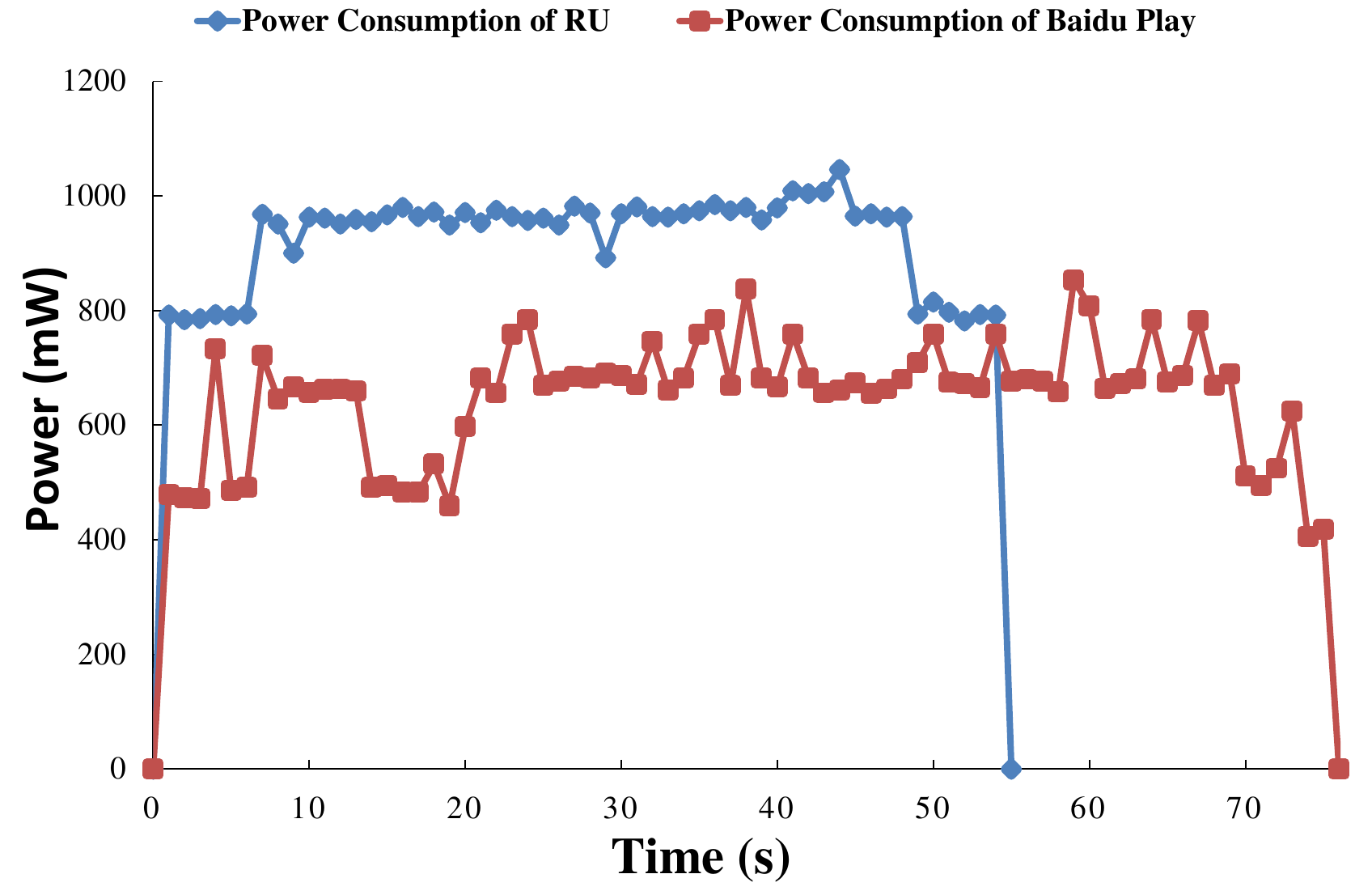}
\caption{Decoding Power Consumption in SAMSUNG I9300: We compare power consumption of our data plan system with Baidu Play(a popular APPs downloaded market software on Android phones in China). Three $AUs$ and one $RU$ are used to download the APP with size 8.67MB in this experiment and record the $RU$'s power consumption. We also record power consumption of $RU$ downloading the same APP from Baidu paly.}
\label{onepower}
\end{center}
\end{figure}

To better illustrate the performance of Prometheus, we present decoding power consumption on RU. AUs just forward the data they download and the power consumption is low. RU's power consumption will influence the system's QoS. If decoder consumes much power of battery, system's QoS is poor and smooth cooperation is difficult. We use server to send encoded symbols and let PowerTutor \cite{powertutor}(A Power Monitor for Android-Based Mobile Platforms) run on SAMSUNG I9300 to record the power consumption of decoder.

Fig.\ref{onepower} shows the change of power consumption during decoding. Average decoding power consumption is 1000mw. The battery capacity of SAMSUNG I9300 is 8.02Wh. Decoder can continue running for $8.02/1 = 8.02$ hours. In practice, decoding power consumption can be accepted and that is worth consuming some power to speed up goodput.

\subsection{Evaluation of the Incentive Scheme}

We assume 20 users are registered to compete and there is only one file download, the file size is 29MB. $\epsilon_{max}$ is uniformly distributed over $[1, \epsilon_{max}]$ and $\gamma=10$. The simulations were run on a Windows machine with 3.00 GHz CPU and 4 GB RAM. Each measurement is
averaged over 100 instances.

Are $\mu$ and $P_{RU}$ strongly affected by the unit costs of AUs. In Table.\ref{jili1}, we can see that $\mu$ and $P_{RU}$ decrease as the unit costs of users become diverse. This is because when the costs become diverse, users with larger costs would have lower chances to join the cooperative transmission.

\begin{table}[h]
\centering
\caption{Impact of $\epsilon_{max}$ on $\mu$ and $P_{RU}$}
\begin{tabular}{|c|c|c|c|c|c|}
\hline
 $\epsilon_{max}$ &1& 2 & 3 & 4& 5 \\
\hline
$\mu$  & 11.9387&10.6457&9.1926&8.3828&6.7315 \\
\hline
$P_{RU}$  & 20.3722&18.3198&17.3723&15.7503&13.3592\\
\hline
\end{tabular}
\label{jili1}
\end{table}

Table.\ref{jili2} shows the impact of $|K|$ on $\mu$ and $P_{RU}$. We fix $\epsilon_{max}=5$ and see $\mu$ and $P_{RU}$ actually increase with $|K|$. The reason is that if more users participate in this system, users with larger unit costs increase the download cost.

\begin{table}[h]
\centering
\caption{$\epsilon_{max}=5$: Impact of $|K|$ on $\mu$ and $P_{RU}$}
\begin{tabular}{|c|c|c|c|c|c|c|}
\hline
 $|K|$&2& 3 & 4 & 5 & 6 &7\\
\hline
$\mu$ &4.274&6.509&7.754&8.591&9.205&9.680 \\
\hline
$P_{RU}$ & 10.706&13.098&14.569&15.616&16.416&17.055\\
\hline
\end{tabular}
\label{jili2}
\end{table}

We also evaluate the impact of $|AU|$ on the $\mu$ and $P_{RU}$ and show the results in Table.\ref{jili3}. Can be obtained from the experimental results: $\mu$ indeed demonstrates diminishing returns when $|AU|$ increases, as will $P_{RU}$.

\begin{table}[h]
\centering
\caption{$\epsilon_{max}=5$: Impact of $|K|$ on $\mu$ and $P_{RU}$}
\begin{tabular}{|c|c|c|c|c|}
\hline
 $|AU|$&5& 10 & 15 & 20\\
\hline
$\mu$  &3.1494&5.232&6.1166&6.7315 \\
\hline
$P_{RU}$  &10.2693&12.0958&13.2726&13.35922\\
\hline
\end{tabular}
\label{jili3}
\end{table}

\section{Conclusion}\label{sec6}

In this paper, we design and evaluate a new data plan sharing system combined with LT codes. LT codes make it possible to download data over lossy wireless links. But the main things here are that the scheme improves the reliability of data transmission and increases the download speed of requesting user significantly. Since smartphones are owned by different individuals, it is reasonable to assume that users are selfish but rational. Hence each user only wants to maximize its own utility, and will not participate in cooperative transmission unless there is sufficient incentive, then we also design an incentive mechanism to encourage participation. We have even implemented cooperative transmission in the release of Android 2.2 or higher, the results of experiments show that the cooperative transmission system is viable. We currently do not take into account the energy efficiency in our data plan system. The power consumption of decoder should be considered.  Moreover, the upload service with cooperative transmission is another important scenario. We will investigate such extensions as part of our future work.


\bibliographystyle{IEEEtran}
\bibliography{mybibISIT}

\end{document}